# Spin-orbit torque-driven magnetization switching and thermal effects studied in Ta\CoFeB\MgO nanowires

R. Lo Conte[1,3], A. Hrabec[2], A. P. Mihai[2†], T. Schulz[1], S.-J. Noh[1], C. H. Marrows[2], T. A. Moore[2], and M. Kläui[1,3]

[1]*Johannes Gutenberg Universität-Mainz, Institut für Physik, Staudinger Weg 7, 55128 Mainz, Deutschland*
[2]*School of Physics and Astronomy, E. C. Stoner Laboratory, University of Leeds, Leeds LS2 9JT, U.K.*
[3]*Graduate School of Excellence Materials Science in Mainz (MAINZ), Staudinger Weg 9, 55128 Mainz, Deutschland*
[†]*now at: Imperial College, South Kensington Campus, London SW7 2AZ, UK*

We demonstrate magnetization switching in out-of-plane magnetized Ta\CoFeB\MgO nanowires by current pulse injection along the nanowires, both with and without a constant and uniform magnetic field collinear to the current direction. We deduce that an effective torque arising from spin-orbit effects in the multilayer drives the switching mechanism. While the generation of a component of the magnetization along the current direction is crucial for the switching to occur, we observe that even without a longitudinal field thermally generated magnetization fluctuations can lead to switching. Analysis using a generalized Néel-Brown model enables key parameters of the thermally induced spin-orbit torques switching process to be estimated, such as the attempt frequency and the effective energy barrier.

Nowadays a large effort is focused on the investigation of novel magnetic materials systems in order to find good candidates for new logic circuits and memory devices. A continuous and growing interest in *spin-orbit torque* (SOT)-driven magnetization dynamics results from the possibility of using this torque in novel spintronic devices based on ultra-fast current-induced domain wall motion[1] or fast current-induced magnetization switching[2]. The origin of these torques in perpendicularly magnetized multilayers with structural inversion asymmetry (SIA) such as Pt\Co\AlOx[3-6] or Ta\CoFe\MgO[7] seems to be spin-orbit effects generated when an electric current is injected through them. One of the two main effects proposed is the Rashba effect[4,5,8,9], which generates an in-plane effective magnetic field perpendicular to both the current-flow and the out-of-plane axis. The second one is the spin Hall effect (SHE)[7,10-13], which generates a pure spin-current diffusing across the heavy metal-ferromagnet interface. Such a spin-current has an in-plane spin polarization, which is able to exert a torque on the magnetic texture present in the ferromagnetic layer. However, the debate about the actual origin of the torques is still open. Systems made of Ta\CoFeB\MgO are of particular interest as such a trilayer is already used as a bottom electrode in existing spintronic devices based on CoFeB\MgO\CoFeB magnetic tunnel junctions (MTJs)[14]. The main issue with such devices is the possible damage of the MTJ due to the injection of current across the tunnel barrier, in order to obtain spin-transfer torque (STT)-driven magnetization switching[15]. However, this risk could be mitigated using a three terminal device where switching is driven by SOTs produced by current flowing in plane[16]. In this case only a small read-out current would flow through the tunnel junction - reducing the risk of structural damage to the insulating barrier from high voltages and currents, and making it possible to scale down the cross-section of the MTJs.

While the switching of structures using SOTs has been investigated previously[6,7,10], the switching was only achieved when an additional longitudinal magnetic field was applied, which is cumbersome for real applications. Additionally, the current is expected to generate Joule heating, so thermal effects, which have not yet been investigated in detail, can be expected to play a role in the SOT-driven magnetization switching.

In this letter, we demonstrate magnetization switching of out-of-plane magnetized Ta\CoFeB\MgO nanowires by current pulse injection along the nanowires. Switching is obtained both with and without a constant and uniform external in-plane magnetic field collinear to the current direction. Current pulses of different time-length and of different polarity are used to obtain switching of the magnetization of the nanowires. The effects of thermal fluctuations in the switching process are analyzed based on a generalized Néel-Brown model and this allows key parameters, such as the attempt frequency and the energy barrier for the switching process, to be determined.

Our sample consists of a Ta(5nm)\Co$_{20}$Fe$_{60}$B$_{20}$(1nm)\MgO(2nm)\Ta(5nm)-multilayer deposited on a thermally oxidized Si-wafer. All the layers have been deposited by sputtering technique (using a Singulus TIMARIS/ROTARIS tool), followed by annealing at 300 $^o$C for 2 hours in order to obtain a large perpendicular magnetic anisotropy. The sample was then patterned into an array of 20 nanowires in a parallel configuration (see Fig. 1) by electron-beam lithography and argon-ion milling. Each nanowire is 1 µm wide and 8 µm long. At both ends of the nanowires there are two magnetic pads, directly connected to two gold pads made in a second optical lithography step. The electrical DC resistance of the device is 130 Ω. As illustrated in Fig. 1, during the measurements the device is electrically connected in series to a pulse generator on one side, and to an oscilloscope on the other. A voltage pulse is applied to the sample and the pulse waveform is measured by the oscilloscope across its 50 Ω-internal resistance ($R_0$). The current flowing through the system is obtained by the measured voltage ($V_o$) across $R_0$. Considering that of the Ta-capping layer only about 3 nm are still conducting due to surface oxidation, we estimate the current density as follows: for 1 V dropping across $R_0$ (corresponding to a total current of 20 mA) we have a current density of 1.1x10$^{11}$ A/m$^2$ flowing through the nanowires of our system. The magnetization state of the nanowires is imaged by differential polar Kerr microscopy. The microscope is also equipped with an in-plane magnetic coil for the generation of an external in-plane magnetic field. During our experiment carried out at T=300 K, we first saturate the magnetic nanowires in the up (+z)- or down (-z)-magnetization state by an external magnet, then we apply a constant and uniform in-plane magnetic field µ$_0$**H$_x$** along the nanowires. After the injection of a current pulse through the device, the magnetization of the nanowires is switched for appropriately chosen field and current density[6,7]. Each magnetization switching event is observed directly using a Kerr microscope, after the current pulse injection (Figs. 1(b) and 1(c)).

In Fig. 2 we show the magnetization switching for 100 ns (closed symbols)- and 100 ms (open symbols)-long current pulses, when a constant and uniform external magnetic field µ$_0$**H$_x$** is applied along the nanowire direction. For both the 100 ns- and 100 ms-long pulses we determine the minimum current amplitude needed to obtain magnetization switching in combination with a fixed external magnetic field value (see Fig. 2(a)). Successful magnetization switching is defined as switching in at least half of the total number of pre-saturated nanowires. From Fig. 2(a) we can see that there is a clear link between the current needed for the switching to happen and the applied in plane field. The lower the in-plane field, the larger is the necessary amplitude of the current pulse. This is qualitatively in agreement with the model of the spin-orbit effective field reported by Emori et al.[7], where $\boldsymbol{H}_{SHE} = \hbar\theta_{SHE}|j_e|/(2\mu_0|e|M_s t_f)(\hat{\boldsymbol{m}} \times (\hat{\boldsymbol{z}} \times \hat{\boldsymbol{j}_e}))$, and $\theta_{SHE}$ is the spin Hall angle, $j_e$ the electron-current density, $M_s$ the saturation magnetization of the ferromagnetic material and $t_f$ the thickness of the ferromagnetic layer. The effective spin-orbit field is also a function of the sign of the current, so that it can work with the external field (leading to a switching event) or against it (keeping the magnetization state in the initial one).

Next, we compare the results obtained for the two different pulse lengths. First of all we find that, the amplitude of the short current pulses required for switching is about 3 times larger than in the case of long pulses, for a fixed value of $H_x$. Furthermore, we find also that for the long pulses we observe a saturation of the current needed for switching for values of $\mu_0 H_x$ lower than 50 mT (see guide lines in Fig. 3 (a)). These observations clearly point to thermal effects playing a crucial role. For large enough current amplitudes, the thermal fluctuations of **M** start to generate an in-plane component of **M**, which becomes comparable to that generated by an external field. When a large enough $M_x$ component is eventually thermally generated in the correct direction, a spin-orbit effective field acts on it yielding a magnetization switching in part of the nanowire and then a complete switching by fast domain wall motion[6]. During the injection of long pulses there is much more time for the thermal induced switching to occur than in the case of the short pulses, increasing the switching probability and thus reducing the necessary in-plane currents. This explains why, even at larger pulses amplitudes, we do not observe this saturation effect for the case of short pulses. Based on that, we check also whether these thermal fluctuations can become sufficiently large to switch the magnetization even without any applied in-plane fields, as described in the last part of this letter. Fig. 2(b) shows the stable magnetization configurations as a function of current sign and field orientation. The table shows that for both $I_p$ and $H_x$ positive or negative the stable magnetization configuration is M+ (up), while for $I_p$ and $H_x$ of opposite sign the stable magnetization state is M- (down). This result confirms the fact that the switching effect is strictly linked to the spin-current generated in the heavy metal when an electric current is injected in plane through the stack[6,7,16,17]. This is also in agreement with the observation in systems made of Ta\CoFe\MgO[7], confirming that the sign of $\theta_{SHE}$ in Ta is negative[16].

Based on our finding of thermally activated switching, we measure the length $\Delta t_{pulse}$ of the current pulse needed to switch the magnetization in our system, using a fixed current density of $5.8 \times 10^{11}$ A/m$^2$ (see Fig. 3). The experiment is carried out with different values of the external magnetic field, measuring the minimum $\Delta t_{pulse}$ needed for a 50% probability of switching. Fig. 3 shows that there is an exponential dependence for the $\Delta t_{pulse}$ on the external magnetic field $\mu_0 H_x$. The lower the applied longitudinal field, the longer is the current pulse needed for switching. For $\mu_0 H_x = 0$ mT we observe magnetization switching in at least 10 of the 20 nanowires of our device, for pulse lengths of 55 ns or more. To obtain the key parameters for the thermally activated switching, we fit our experimental data with a generalized Néel-Brown model[18]. This model describes the role of the thermal fluctuations in magnetization switching processes, in the presence of a spin-transfer torque. In our experiment we have a similar situation, with an electric current flowing in the system and a spin-current pumped in the magnetic layer due to the SHE, giving a non-equilibrium condition. The equation that describes our experiment is the following[18]:

$$\Delta t_{pulse} = 0.7 f_0^{-1} \exp\left(\frac{E_0^*}{k_B T}\left(1 - \frac{H_z}{H_{sw}^0}\right)^{1.5}\right)$$

where $f_0$ is the attempt frequency, $E_0^*$ is the energy barrier at the given current, $\mu_0 H^\circ_{sw}$ is the switching field at 0 K, and $\mu_0 H_z$ is the field along the easy axis. In our case the field along the z-direction is the spin-orbit effective field $H_z = \frac{\hbar\theta_{SHE}|j_e|}{2\mu_0|e|M_s t_f} m_x = \frac{\hbar\theta_{SHE}|j_e|}{2\mu_0|e|M_s t_f} \frac{H_x}{H_x^{sat}}$, where $\mu_0 H_x^{sat}$ is the longitudinal field needed for aligning all the magnetization in the x-direction ($\approx$ 400 mT in our system). As can be seen from Fig. 3, the model fits the data well. From the fit we obtain an attempt frequency $f_0 \approx$ 660 MHz and an effective energy barrier $E_0^* \approx$ 190 meV. This is a lower boundary of the actual energy barrier $E_0$, due to the fact that the electric current and the spin-current flowing in the magnetic layer reduce its effective value. From measurements of the sample resistance as a function of the temperature and of the current density during current injection[19], we deduce a quasi-static temperature of about 650 K in the nanowires during the current pulse injection at this high current density.

From the fitting curve we find an off-set of 19 ns in order to be able to fit our data. We can explain the presence of such an off-set by to two main reasons. The first reason is the shape of the pulses we are able to apply with our set-up. As it can be seen in the in-set in figure 1(a), we have a rise- and a fall-time of about 5 ns each. This means that, the current density is actually at its top value for at least 10 ns less than the nominal duration of the pulse. Secondly, we can assume that a nanowire is switched when the magnetization is inverted in at least the 90% of it. For a switching mechanism based on domain nucleation followed by domain wall motion[7], part of the minimum time obtained from the fitting curve could also be due to the fact that the domain walls need some time to move to the end of the nanowires. The calculated domain wall speed is in the order of 200-300 m/s, which is in line with the velocities measured for similar materials stacks[3-5].

Finally, we note that this thermally activated switching is not unidirectional, since in absence of the magnetic field the spin-orbit torque would destabilize the magnetization configuration every time a current pulse is injected in the system. This means we can obtain switching in both directions using the same current pulse polarity. For applications this means that one would need to read after writing to check that the writing has been successful.

In summary, we have demonstrated SOT-driven magnetization switching in nanowires made of Ta\CoFeB\MgO. The switching was observed to depend on the pulse length for 100 ns- and 100 ms-long current pulses in the current density range of $10^{11}$ A/m$^2$ due to thermal effects. Pure current-induced magnetization switching in absence of external magnetic field applied was obtained for 55 ns-long pulses at $5.8 \times 10^{11}$ A/m$^2$. Our experimental data are reproduced by employing the generalized Néel-Brown model, enabling the attempt frequency and the effective energy barrier to be extracted. While this approach allows us to switch magnetization even without any fields, it requires a subsequent read-out to confirm switching for the use in devices.

**Acknowledgments**

We acknowledge support by the Graduate School of Excellence Materials Science in Mainz (MAINZ) GSC 266, Staudinger Weg 9, 55128, Germany; the EU (IFOX, NMP3-LA-2012 246102; MASPIC, ERC-2007-StG 208162; WALL, FP7-PEOPLE-2013-ITN 608031), and the Research Center of Innovative and Emerging Materials. This work was also supported by EPSRC, U.K. (Grant Nos. EP/I011668/1, EP/G005176/1, EP/K003127/1) and the Alexander von Humboldt Foundation CONNECT program.

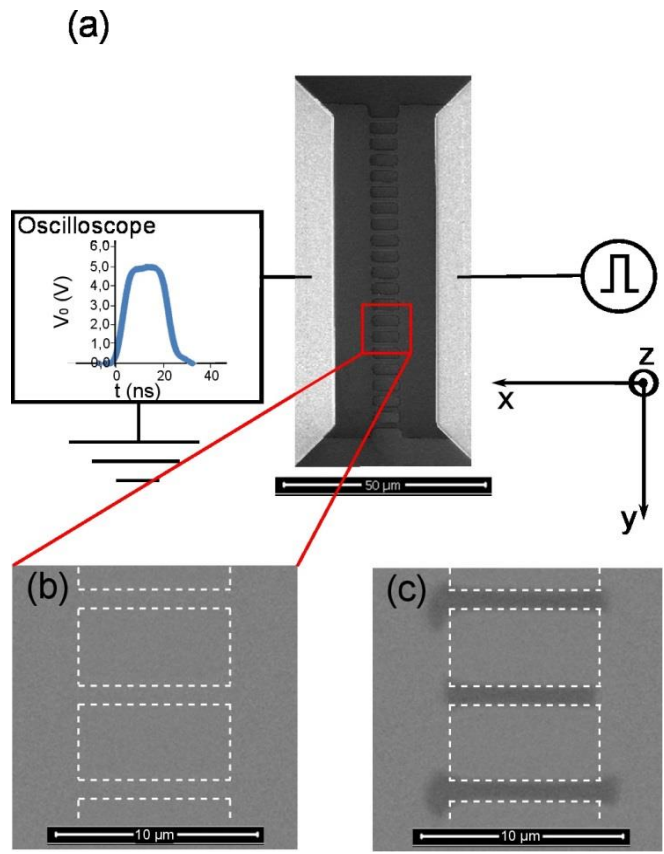

FIG. 1. (a) Schematic of the experimental set-up for current pulse injection, including an SEM micrograph of the sample used during the experiment. The inset shows the shape of one of the voltage pulses applied to the device, measured with the oscilloscope (across the 50 Ω internal resistance). (b) Differential Kerr microscopy image of the initialized nanowires with the magnetization pointing down (-z) everywhere. (c) Differential Kerr microscopy image of the same nanowires in (b), after their magnetization has been switched up (+z) by a current pulse in the presence of an in-plane magnetic field collinear with the current-flow.

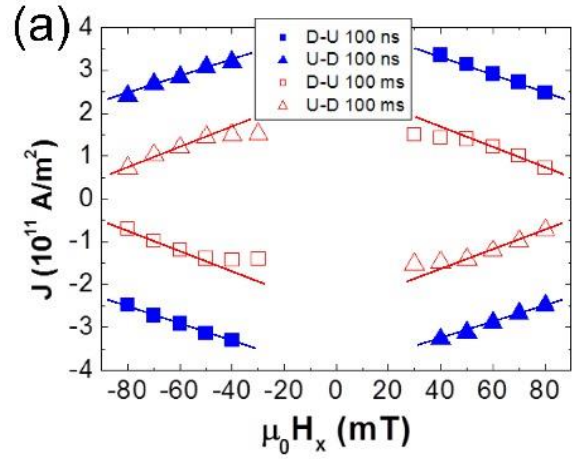

FIG. 2. (a) Graph showing the minimum current density (**J**) needed to switch the magnetization in at least half of the nanowires (10 of 20) for a fixed longitudinal in-plane magnetic field ($\mu_0 H_x$). The closed (open) symbols show the switching events due to 100 ns- (100 ms-) long current pulses. The squares (triangles) show the Down → Up (Up → Down) switching events. The lines are a guide to the eye, highlighting the saturation of the required switching current density at low magnetic fields in the case of long pulses. (b) The table shows the stable magnetization state for the four different field-current combinations: M+ (+z-direction), M- (-z-direction).

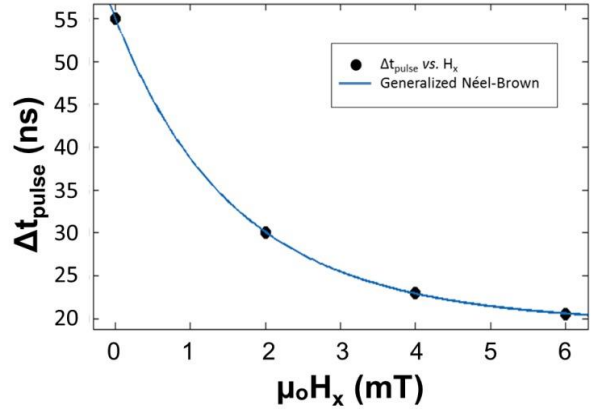

FIG. 3. Graph showing the minimum length of the current pulse needed in order to switch the magnetization in at least 10 of the 20 nanowires of the sample, as function of the longitudinal magnetic field. The pulse current density is $5.8 \times 10^{11}$ A/m$^2$. We observe successful magnetization switching at zero field, for current pulses of at least 55 ns in length. The fitting curve is based on the generalized Néel-Brown model for thermal-assisted magnetization switching.